\documentclass[prd,twocolumn,preprintnumbers,superscriptaddress,showpacs]{revtex4-1}
\usepackage[colorlinks=true, pdfstartview=FitV, linkcolor=red, citecolor=blue, urlcolor=black, pdftitle={},pdfauthor={Tomoya Hayata},pdfsubject={}, pdfkeywords={}]{hyperref}

\usepackage{graphicx}% Include figure files
\usepackage{dcolumn}% Align table columns on decimal point
\usepackage{bm}% bold math
\usepackage{subfigure}
\usepackage{slashed}
\usepackage{comment}
\usepackage{booktabs}
\usepackage{amssymb}
\usepackage{amsmath}
\usepackage{color}
\newcommand{\red}[1]{{\color{red} #1} }

\usepackage[normalem]{ulem}

\begin{document}
\title{Non-equilibrium Chiral Magnetic/Vortical Effects in Viscous Fluids}
\author{Yoshimasa Hidaka}
\affiliation{
	Nishina Center,
	RIKEN, Wako, Saitama 351-0198, Japan.\\
}
\affiliation{
	iTHEMS Program, RIKEN, Wako, Saitama 351-0198, Japan.
}
\author{Di-Lun Yang}
\affiliation{
	Nishina Center,
	RIKEN, Wako, Saitama 351-0198, Japan.\\
}
%\date{\today}
\begin{abstract}
We utilize the chiral kinetic theory in a relaxation-time  approximation to investigate the nonlinear anomalous responses of chiral fluids with viscous effects. Unlike the cases in equilibrium, it is found that the chiral magnetic effect  and chiral vortical effect  are modified by the shear and bulk strengths. Particularly, the shear strength could result in charged Hall currents for chiral magnetic  and chiral vortical effects, which propagate perpendicular to applied magnetic fields and vorticity. These quantum corrections stemming from side jumps and anomalies are dissipative and pertinent to interactions.
Although the non-equilibrium effects upon charge currents are dissipative, the second law of thermodynamics is still satisfied.      
\end{abstract}

%\keywords{Chiral Kinetic Theory, Chiral Anomalies, Weyl Fermions, Chiral Fluids}
\maketitle

\section{Introduction}
The anomalous transport for Weyl fermions related to quantum anomalies such as chiral magnetic/vortical effects (CME/CVE), from which charged currents are induced by magnetic/vortical fields, has recently aroused great interest in the studies of relativistic heavy ion collisions (HIC) and Weyl semimetals  \cite{Vilenkin:1979ui, Kharzeev:2007jp,Fukushima:2008xe,Li:2014bha}. Such effects and relevant phenomena associated with chiral imbalance have been investigated from various approaches including field-theory calculations based on Kubo formula \cite{Fukushima:2008xe,Kharzeev:2009pj,Landsteiner:2011cp}, kinetic theory \cite{Gao:2012ix, Son:2012wh, Stephanov:2012ki, Son:2012zy,Chen:2012ca,Manuel:2013zaa,
	Manuel:2014dza,Kharzeev:2016sut, Huang:2017tsq,Gao:2017gfq,Ebihara:2017suq}, 
relativistic hydrodynamics \cite{Son:2009tf,Neiman:2010zi,Sadofyev:2010pr,Pu:2010as,Kharzeev:2011ds},
lattice simulations \cite{Abramczyk:2009gb,Buividovich:2009wi,Buividovich:2009zzb,Buividovich:2010tn,Yamamoto:2011gk,Mueller:2016ven,Mace:2016shq}, and
gauge/gravity duality \cite{Erdmenger2009,Torabian2009a,Banerjee2011,Landsteiner2011}. Particularly, recent progress in chiral kinetic theory (CKT) with the manifestation of Lorenz symmetry related to side jumps and the incorporation of collisions has improved our understandings on anomalous transport out of equilibrium \cite{Chen:2014cla,Chen:2015gta,Hidaka:2016yjf}.

It is generally believed that CME in equilibrium is protected by the chiral anomaly and unaffected by interactions. On the other hand, CVE in equilibrium could be protected by mixed-axial-gravitational anomaly with only background fields \cite{Landsteiner2011,Landsteiner:2011cp, Jensen:2012kj}, whereas the modification from interactions could emerge in the presence of dynamical gauge fields \cite{Hou2012,Golkar:2012kb,Jensen:2013vta}. See Refs.~\cite{Chowdhury:2015pba,Golkar:2015oxw,Chowdhury:2016cmh} for some following works. In addition, there are studies for the axial currents induced by electric fields and chiral imbalance such as the chiral electric separation effect (CESE) or chiral Hall effect (CHE), which are pertinent to interactions \cite{Huang:2013iia,Pu:2014cwa,Pu:2014fva}, while these effects are not directly connected to the helicity of Weyl fermions and quantum anomalies. 
Nonetheless, an interplay between the chiral anomaly, magnetization current stemming from side jumps, and magnetic-moment coupling could result in non-equilibrium corrections involving interactions upon anomalous transport. For example, the alternative-current (AC) conductivity for CME with nonzero frequency is modified in the presence of time-dependent magnetic fields \cite{Kharzeev:2009pj,Satow:2014lva,Kharzeev:2016sut}. Also, the nonlinear responses on anomalous transport triggered by fluctuations near local equilibrium in inviscid chiral fluids have been analyzed by some of the authors in this paper \cite{Hidaka:2017auj}, where novel anomalous Hall currents led by electric fields and temperature/chemical-potential gradients are found \footnote{See Refs.~\cite{Chen:2016xtg,Gorbar:2016qfh} for related studies in open systems, where the energy-momentum conservation led by hydrodynamics is not considered.}. However, the non-equilibrium corrections on charge currents from magnetic fields or vorticity may appear in the viscous case \footnote{The viscous effects in the absence of background fields are incorporated in the study presented in Ref.~\cite{Gorbar:2017toh}, while the authors focus on the hydrodynamic dispersion relation modified by vorticity and the charge current is not computed therein.}. Experimentally, in HIC, due to the viscous corrections in the quark gluon plasma, the non-equilibrium effects upon the CME conductivity may impact the charge separation associated with the signal for CME. Moreover, based on recent observations of hydrodynamic transport in graphene \cite{Crossno1058,PhysRevLett.116.136802}, there exists mounting interest in the possible realization of chiral fluids in Weyl semimetals (see e.g., Refs.~\cite{Lucas:2016omy, Gorbar:2017vph} for relevant studies in theory). The theoretical investigation of relativistic viscous hydrodynamics of chiral fluids is thus imperative.

In this paper, we follow the approach in Ref.~\cite{Hidaka:2017auj} to further study the viscous corrections on anomalous transport contributed by the aforementioned quantum effects associated with the helicity of Weyl fermions. It is found that the CME and CVE conductivities of charge currents are modified by shear and bulk strengths. Nevertheless, the non-equilibrium corrections on the energy-density currents should vanish according to the matching condition in the classical relaxation-time approximation (RTA). Although the non-equilibrium quantum corrections are dissipative, the second law of thermodynamics is still satisfied.  

\section{Wigner functions and chiral kinetic theory}
We begin with a brief introduction to the Wigner-function formalism of CKT which will be exploited to study the non-equilibrium transport.
As derived in Ref.~\cite{Hidaka:2016yjf} by solving Dirac equations up to $\mathcal{O}(\hbar)$ from the Wigner-function approach, the perturbative solution for the lesser propagators of right-handed Weyl fermions reads   
\begin{eqnarray}\label{Wigner_S}\nonumber
\grave{S}^{<\mu}(q,X)
&=&2\pi\bar{\epsilon}(q\cdot n)\Big(q^{\mu}\delta(q^2)
+\hbar\delta(q^2)S_{(n)}^{\mu\nu}\mathcal{D}_{\nu}
\\
&&
+\hbar\epsilon^{\mu\nu\alpha\beta}q_{\nu}F_{\alpha\beta}\frac{\partial\delta(q^2)}{2\partial q^2}
\Big)f^{(n)}_q,
\end{eqnarray}
where $\bar{\epsilon}(q\cdot n)$ represents the sign of $q\cdot n$ and
\begin{eqnarray}
S^{\mu\nu}_{(n)}=\frac{\epsilon^{\mu\nu\alpha\beta}}{2(q\cdot n)}q_{\alpha}n_{\beta} 
\label{S_n_1}
\end{eqnarray}
corresponds to the spin tensor depending on a frame vector $n^{\mu}$~\cite{Chen:2015gta}. 
The frame vector can be understood as the zeroth component of a vierbein for the local transformation of $\sigma^{a}=(I,\sigma^i)$ introduced on the local tangent space to $\sigma^{\mu}(X)$ depending on the global spacetime coordinates, where $\sigma^i$ represent Pauli matrices for $i=1,2,3$, such that $\sigma^{\mu}=e^{\mu}_a\sigma^{a}$. That is, we define $n^{\mu}\equiv e^{\mu}_0$ 
with $\mu$ being the spacetime indices (See Appendix~\ref{sec:WignerFunction} for details of the choice of the  frame vector). In flat spacetime, we may set $e^{\mu}_{a}(X)=\delta^{\mu}_a$ such that the spin connection vanishes.
Now, the global spacetime coordinate transformation corresponds to the frame transformation $n^{\mu}\rightarrow n'^{\mu}$.

Here we denote $\mathcal{D}_{\beta}f^{(n)}_q=\Delta_{\beta}f^{(n)}_q-\mathcal{C}_{\beta}$, where $\Delta_{\mu}=\partial_{\mu}+F_{\nu\mu}\partial^{\nu}_{q}$,  $\mathcal{C}_{\beta}=\Sigma_{\beta}^<\bar{f}^{(n)}_q-\Sigma_{\beta}^>f^{(n)}_q$ with $\Sigma_{\beta}^{<(>)}$ being less/greater self-energies and $f^{(n)}_q$ and $\bar{f}^{(n)}_q=1-f^{(n)}_q$ being the distribution functions of incoming and outgoing particles, respectively. The second term in Eq.~\eqref{Wigner_S} associated with $S^{\mu\nu}_{(n)}$ as the side-jump term only contributes for the non-equilibrium cases or for a rotating system in global equilibrium, which results in magnetization currents and CVE. On the other hand, the third term in Eq.~\eqref{Wigner_S} yields CME in equilibrium. 

Given the Wigner functions, one can directly evaluate the charge current and energy-momentum tensor through
\begin{eqnarray}\label{def_J}\nonumber
&&T^{\mu\nu}=\int \frac{d^4q}{(2\pi)^4}\big(q^{\mu}\grave{S}^{<\nu}+q^{\nu}\grave{S}^{<\mu}\big),
\\
&&J^{\mu}=2\int \frac{d^4q}{(2\pi)^4}\grave{S}^{<\mu}.
\end{eqnarray} 
The distribution function in $\grave{S}^{<\mu}$ has to be solved from the CKT led by the Dirac equation. In light of the study in Ref.~\cite{Hidaka:2017auj}, we focus on the fluctuations slightly away from the local equilibrium distribution function defined in a comoving frame $n^{\mu}=u^{\mu}$ with $u^{\mu}$ being a fluid velocity.

	For an arbitrary frame $n^{\mu}$, the CKT takes the form~\cite{Hidaka:2017auj},
	\begin{eqnarray}\label{CKT}
	\delta\Bigl(q^{2}-\hbar \frac{B\cdot q}{q\cdot n}\Bigr)\Box(q,X)f^{(n)}_q=\delta\Bigl(q^{2}-\hbar \frac{B\cdot q}{q\cdot n}\Bigr)\mathcal{C}_\text{full}, 
	\end{eqnarray}
	where 
	\begin{eqnarray}\nonumber
	\Box(q,X)&=&\Big[
	q\cdot\Delta+\hbar\frac{S_{(n)}^{\mu\nu}E_{\mu}}{(q\cdot n)}\Delta_{\nu}
	+\hbar S_{(n)}^{\mu\nu}(\partial_{\mu}F_{\rho \nu})\partial^{\rho}_{q}
	\\
	&&+\hbar(\partial_{\mu}S^{\mu\nu}_{(n)})\Delta_{\nu}\Big],
	\end{eqnarray}
	and 
	\begin{eqnarray}\label{C_full}\nonumber
	\mathcal{C}_\text{full}&=&\Bigg(q^{\mu}+\hbar\frac{S_{(n)}^{\nu\mu}E_{\nu}}{(q\cdot n)}+\hbar\big(\partial_{\rho}S^{\rho\mu}_{(n)}\big)\Bigg)\tilde{\mathcal{C}}_{\mu},
	\\
	\tilde{\mathcal{C}}^{\mu}&=&\mathcal{C}^{\mu}+\hbar\frac{\epsilon^{\mu\nu\alpha\beta}n_{\nu}}{2q\cdot n} 
	\big(\bar{f}^{(n)}_q\Delta^>_{\alpha}\Sigma^{<}_{\beta}-f^{(n)}_q\Delta^<_{\alpha}\Sigma^{>}_{\beta}\big)
	.
	\end{eqnarray}
	Here, we define $B^{\mu}$ and $E_{\mu}$ by decomposing the field strength into $F_{\alpha\beta}=-\epsilon_{\mu\nu\alpha\beta}B^{\mu}n^{\nu}+n_{\beta}E_{\alpha}-n_{\alpha}E_{\beta}$. We will then work in $n^{\mu}=u^{\mu}$.
The collision term depends on details of the system, so that it is not universal. Here, 
for simplicity and generality to be applicable for various systems including e.g., HIC or Weyl semimetals and also to make comparisons with previous studies \cite{Chen:2016xtg,Gorbar:2016qfh} on equal footing, we apply the RTA for the collisional kernel in order to capture the qualitative features based on the symmetric properties of collisions. We thus approximate
\begin{eqnarray}\label{RT_approx}
\mathcal{C}_\text{full}\simeq-\frac{1}{\tau_R}\Big(q\cdot u+\hbar \frac{q^{\mu}\mathcal{A}_{\mu}}{(q\cdot u)^2}\Big)\delta f_q,
\end{eqnarray}
where $\mathcal{A}_{\mu}$ denotes an operator acting on $\delta f_q$, which represents possible quantum corrections with $\mathcal{O}(\partial)$ that depends on the details of collision terms. $\tau_R$ is the relaxation time charactering the inverse strength of interactions, which will be treated as a constant \red{\footnote{In fact, naively taking $\tilde{\mathcal{C}}_{\mu}=-\tau_R^{-1}u_{\mu}\delta f_q$ in Eq.~\eqref{C_full}, one finds $\mathcal{A}_{\mu}=-\omega_{\mu}$. Nonetheless, the $\Sigma^{<(>)}_{\mu}$ in $\mathcal{C}_{\mu}$ actually contains $\hbar$ corrections stemming from Wigner functions, which depend on the details of realistic collisions such as 2 to 2 scattering or the interactions with media. Such contributions hence make $\mathcal{A}_{\mu}$ undetermined without specifying a particular scattering process.}}. Note that here $\mathcal{A}_{\mu}$ is also frame dependent. The transformation of $\mathcal{A}_{\mu}$ in different frames should preserve the frame independence (Lorentz invariance) of the CKT with the RTA. We introduce such a frame transformation upon $\mathcal{A}_{\mu}$ in Appendix~\ref{sec:fRTA}.

We will then follow the computations in Ref.~\cite{Hidaka:2017auj} to perturbatively solve for the non-equilibrium distribution function $\delta f_q=f^{(u)}_q-f^\text{eq}_q$ incorporating viscous corrections from the CKT in Eq.~\eqref{CKT} based on the $\hbar$ and derivatives expansions. Here the local equilibrium distribution function takes the form $f^\text{eq}_q=(e^{g}+1)^{-1}$ with
$g=(q\cdot u-\mu+\hbar(\omega\cdot q)(2q\cdot u)^{-1})/T$ for $T$ and $\mu$ being the local temperature and chemical potential, respectively. Also, the vorticity $\omega^{\mu}$ is defined as $\omega^{\mu}\equiv \epsilon^{\mu\nu\alpha\beta}u_{\nu}\big(\partial_{\alpha}u_{\beta}\big)/2$. 

\textit{Hydrodynamics and matching conditions}.---
Furthermore, following the charge and energy-momentum conservation with the chiral anomaly, we should also implement the anomalous hydrodynamic EOM led by
\begin{eqnarray}\label{cons_eq}
\partial_{\mu}J^{\mu}=-\frac{\hbar}{4\pi^2}E_{\mu}B^{\mu}
,\quad \partial_{\mu}T^{\mu\nu}=F^{\nu\mu}J_{\mu}.
\end{eqnarray}
These two equations provide the physical constrains for CKT, which dictate the dynamics of thermodynamic parameters $T$, $\mu$, and $u^{\mu}$ in $f^{\text{eq}}$.
However, by utilizing the equation $\Delta_{\mu}\grave{S}^{<\mu}=\Sigma^<_{\mu}\grave{S}^{>\mu}-\Sigma^>_{\mu}\grave{S}^{<\mu}$ as the origin of the CKT, it is shown in Ref.~\cite{Hidaka:2017auj} that the divergence of currents manifests the chiral anomaly,
\begin{eqnarray}\label{div_Jmu}\nonumber
\partial_{\mu}J^{\mu}&=&-\frac{\hbar}{4\pi^2}E_{\mu}B^{\mu}+2\int_q\Big[\delta(q^2)q^{\mu}
\\
&&+\hbar\epsilon^{\mu\nu\alpha\beta}F_{\alpha\beta}\frac{\partial_{q\nu}\delta(q^2)}{4}\Big]\tilde{\mathcal{C}}_{\mu},
\end{eqnarray}
where we denotes $\int_q=\int\frac{d^4q}{(2\pi)^3}\bar{\epsilon}(q\cdot n)$.  
Moreover, performing similar computations as in the case for $\partial_{\mu}J^{\mu}$, the divergence of the energy-momentum gives rise to
\begin{eqnarray}\label{div_Tmunu}\nonumber
\partial_{\mu}T^{\mu\nu}&=&F^{\nu\mu}J_{\mu}+2\int_q\delta(q^2)\Big[q^{\nu}q^{\mu}
+\frac{\hbar\epsilon^{\sigma\mu\alpha\beta}}{4}
\\
&&\times
\Big(\delta^{\nu}_{\sigma}\left(q_{\beta}\partial_{\alpha}+F_{\alpha\beta}\right)
+q^{\nu}F_{\alpha\beta}\partial_{q\sigma}\Big)
\Big]\tilde{\mathcal{C}}_{\mu}.
\end{eqnarray} 
%(\blue{See the Appendix~\ref{sec:EM}} for the detailed derivation.)
The detailed derivation is shown in Appendix~\ref{sec:EM}. For collisions in practical systems obeying charge and energy-momentum conservation, the collisional terms in Eqs.~\eqref{div_Jmu} and \eqref{div_Tmunu} should automatically vanish.
In contrast, in the RTA, the charge and energy-momentum conservations give matching conditions. In particular, when $\mathcal{A}_{\mu}=0$, we find that Eqs.~\eqref{div_Jmu} and \eqref{div_Tmunu} become
\begin{eqnarray}\label{cons_eq_tauR}\nonumber
\partial_{\mu}J^{\mu}&=&-\frac{\hbar}{4\pi^2}E_{\mu}B^{\mu}-\frac{u_{\mu}\delta J^{\mu}}{\tau_R}
,
\\
\partial_{\mu}T^{\mu\nu}&=&F^{\nu\mu}J_{\mu}-\frac{u_{\mu}\delta T^{\mu\nu}}{\tau_R}.
\end{eqnarray}
In this case,  the matching conditions turns to the standard ones, $u_{\mu}\delta J^{\mu}=u_{\mu}\delta T^{\mu\nu}=0$. 
The vanishing non-equilibrium modifications upon charge density and energy-density current consequently allow us to consistently define the temperature and chemical potential in equilibrium as alternatively indicated in Ref. \cite{Gorbar:2017toh}. 
Taking other approximations could result in distinct conserved quantities without consistent physical interpretations.

\section{Non-equilibrium responses and charge currents}
Following Ref.~\cite{Hidaka:2017auj}, the perturbative solution for the non-equilibrium distribution function is given by
\begin{eqnarray}
\delta f_q=
-\frac{\tau_R}{(q\cdot u)}\Big(1-\hbar \frac{q^{\mu}\mathcal{A}_{\mu}}{(q\cdot u)}\Big)\Box f_q^{\text{eq}},
\end{eqnarray}
where $\mathcal{A}_{\mu}$ is taken to be a constant here.
We may make the decomposition,
$\delta f_q=\delta f^{(c)}_q+\delta f^{(Q)}_q$, 
where the superindices $(c)$ and $(Q)$ correspond to the classical and quantum corrections, respectively. We then further separate the part for ideal fluids, and the viscous correction, $\delta f^{(c/Q)}=\delta_I f^{(c/Q)}_q+\delta_v f^{(c/Q)}_q$, where the subindices $I$ and $v$ denote the inviscid and viscous parts. The explicit expression of $\delta_I f_q$ can be found in Ref.~\cite{Hidaka:2017auj}. In light of Ref.~\cite{Hidaka:2017auj}, we decompose the quantum corrections of the non-equilibrium distribution function into three parts as $\delta_v f^{(Q)}_q=\delta_v f^{\mathcal{K}}_q+\delta_v f^{\mathcal{H}}_q+\delta_v f^{\mathcal{C}}_q$, where $\delta_v f^{\mathcal{K}}_q$ is led by the perturbative solution out of equilibrium solved from CKT and $\delta_v f^{\mathcal{H}}_q$ is attributed to the $\hbar$ corrections of the temporal derivatives ($u\cdot\partial$) upon $T$, $\bar{\mu}\equiv\mu/T$, and $u^{\mu}$ from hydrodynamic EOM obtained from Eq.~\eqref{cons_eq}. Finally, $\delta_v f^{\mathcal{C}}_q$ comes from the $\hbar$ corrections in the collisional kernel, while this term depends on the assumption of $\mathcal{A}_{\mu}$ in the RT approximation, which does not play a significant role in our analysis. In Ref.~\cite{Hidaka:2017auj} for an inviscid case, $\mathcal{A}_{\mu}$ is treated as a constant. We will follow the same convention in the computation of viscous corrections for consistency although $\mathcal{A}_{\mu}$ can be an operator governed by the frame transformation. On the other hand, as discussed previously, we may set $\mathcal{A}^{\mu}=0$ in the co-moving frame for self-consistency to introduce local equilibrium thermodynamical parameters suggested by the matching conditions.

For convenience, hereafter we denote $q_0\equiv q\cdot u$ and $D\equiv u\cdot\partial$. By implementing CKT with the RT approximation, in the local rest frame, we find
\begin{eqnarray}
\delta_v f^{(c)}_q=-\frac{\tau_R}{q_0}q^{\mu}q^{\nu}\pi_{\mu\nu}\partial_{q_0}f^{(0)}_q
\end{eqnarray} 
with $f^{(0)}_q=1/(e^{(q_0-\mu)/T}+1)$ and 
\begin{widetext}
\begin{eqnarray}\nonumber
\delta_v f^{\mathcal{K}}_q&=&
\frac{\hbar\tau_R}{2q_0}\Bigg[\frac{B^{\mu}}{q_0}\left(\frac{2q_{\mu}\theta}{3}-\pi_{\mu\nu}q^{\nu}\right)
-2\omega^{\mu}\pi_{\mu\nu}q^{\nu}
+\frac{(q\cdot\omega)}{q_0^2}\left(
q^{\mu}q^{\nu}\pi_{\mu\nu}+\frac{q^2_{\perp}}{3}\theta\right)(1-q_0\partial_{q_0})
\\
&&-\frac{2q\cdot\omega}{3}\theta
-\frac{\epsilon^{\mu\nu\alpha\beta}}{q_0}u_{\mu}q_{\alpha}q^{\rho}\pi_{\nu\rho}
\Big(T\partial_{\beta}\bar{\mu}+\frac{q_0\partial_{\beta}T}{T}+q_0Du_{\beta}\Big)\Bigg]\partial_{q_0}f^{(0)}_q
,
\end{eqnarray}
\end{widetext}
where we define $P^{\mu\nu}\equiv\eta^{\mu\nu}-u^{\mu}u^{\nu}$ as a projection operator
with the Minkowski metric $\eta^{\mu\nu}=\text{diag}(1,-1,-1,-1)$, $\theta\equiv\partial\cdot u$ as the bulk strength, 
$\pi^{\mu\nu}\equiv P^{\mu}_{\rho}P^{\nu}_{\sigma}(\partial^{\rho}u^{\sigma}
+\partial^{\sigma}u^{\rho}-2\eta^{\rho\sigma}\theta/3)/2$ as the shear strength, and
$V_{\perp}^{\mu}\equiv P^{\mu}_{\nu}V^{\nu}$ as the transverse component of an arbitrary vector $V^{\mu}$.

Next, we shall consider the viscous corrections on the hydrodynamic EOM. By solving Eq.~\eqref{cons_eq} with the current and energy-momentum tensor in local equilibrium, it is found that $D T=-\theta/3+\mathcal{O}(\hbar)$ and $D\bar{\mu}=\mathcal{O}(\hbar)$ with $\bar{\mu}\equiv\mu/T$, while the viscous correction does not lead to $\hbar$ corrections for $D T$ and $D \bar{\mu}$. We also include their contributions for the computation of currents. 
In addition, we find that the viscous correction gives rise to the $\hbar$ correction upon $ D_v u^{\mu}_{\perp}$ and accordingly the hydrodynamic EOM results in 
\begin{eqnarray}\nonumber
\delta_v f^{\mathcal{H}}_q&=&-\hbar\tau_RD_vu^{\mu}_{\perp}
\\\nonumber
&=&
-\hbar\tau_R q_{\mu}\Bigg[
\tilde{U}_{\omega}T\omega_{\nu}\left(\frac{P^{\mu\nu}\theta}{6}-\pi^{\mu\nu} \right)
\\&&
+\tilde{U}_BB_{\nu}\pi^{\mu\nu}
\Bigg]
\partial_{q_0}f^{(0)}_q.
\end{eqnarray}
where $\tilde{U}_B=-T^2(3\bar{\mu}^2+\pi^2)/(48p\pi^2)$ and $\tilde{U}_{\omega}=T^2\bar{\mu} \left(\bar{\mu}^2+\pi ^2\right)/(12p\pi^2)$
with $p$ being pressure. On the other hand, the $\hbar$ corrections in collisions give
\begin{eqnarray}
\delta_v f^{\mathcal{C}}_q=
\frac{\hbar\tau_R}{q_0^4}\left(q^{\mu}q^{\nu}\pi_{\mu\nu}+\frac{q_{\perp}^2}{3}\theta\right)
(q\cdot\mathcal{A})
\partial_{q_0}f^{(0)}_q.
\end{eqnarray}

From Eqs.~\eqref{Wigner_S} and \eqref{def_J}, the quantum corrections of the non-equilibrium current reads 
\begin{eqnarray}\label{JvQ}\nonumber
\delta J_{Q}^{\mu}&=&2\hbar\int\frac{d^4q}{(2\pi)^3}\bar{\epsilon}(q\cdot u)\delta(q^2)\Bigg[ q^{\mu}\delta f^{(Q)}_q
\\&&
+\Big(S^{\mu\nu}_{(u)}\Delta_{\nu}
-\frac{\epsilon^{\mu\nu\alpha\beta}}{4}F_{\alpha\beta}\partial_{q\nu}\Big)
\delta f^{(c)}_q
\Bigg].
\end{eqnarray}
By inserting $\delta f^{(Q)}_q$ and $\delta f^{(c)}_q$ into Eq.~\eqref{JvQ}, we obtain the quantum correction upon the charge current, $\delta J^{\mu}_{Q\perp}=\delta J^{\mu}_{IQ\perp}+\delta J^{\mu}_{vQ\perp}$, where $\delta J^{\mu}_{IQ\perp}$ as the part for ideal fluids is shown in Ref.~\cite{Hidaka:2017auj}. The viscous part takes the form 
\begin{eqnarray}\label{non_cond}
\delta_v J^{\mu}_{Q\perp}=
\hbar\left(\delta{\sigma}^{\mu\nu}_BB_{\nu}+\delta{\sigma}^{\mu\nu}_{\omega}\omega_{\nu}+\delta\sigma^{\mu\nu}_{\mathcal{A}}\mathcal{A}_{\nu}
\right).
\end{eqnarray}
When not applying the hydrodynamic EOM, only $\delta_v f^{\mathcal{K}}_q$ contributes and one finds
\begin{eqnarray}\nonumber
\delta{\sigma}^{\mu\nu}_B&=&\frac{\tau_R\mu}{4\pi^2}
\Big(\frac{4}{9}\theta P^{\mu\nu}
-\pi^{\mu\nu}\Big),
\end{eqnarray}
\begin{eqnarray}
\nonumber
\delta{\sigma}^{\mu\nu}_{\omega}&=&-\frac{\tau_RI_1T^2}{36\pi^2}
\Big(\frac{17P^{\mu\nu}\theta}{3}
+20\pi^{\mu\nu}
\Big),
\end{eqnarray}
and
\begin{eqnarray}
\delta{\sigma}^{\mu\nu}_{\mathcal{A}}=-\frac{\tau_R\mu}{18\pi^2}\Big(\frac{5P^{\mu\nu}\theta}{3}+2\pi^{\mu\nu}\Big),
\end{eqnarray}
where $I_1=\bar{\mu}^{2}+{\pi^{2}}/{3}$.
The results suggest that the viscous corrections upon CME and CVE conductivities should exist even for an open system in which the back-reaction on environments is neglected and the energy-momentum conservation is violated such as the case in Weyl semimetals when the scattering between quasi-particles and impurities dominate the interactions among quasi-particles. 

Now, for the right-handed chiral fluid as a closed systems with energy-momentum conservation, by implementing the hydrodynamic EOM, the coefficients become
\begin{eqnarray}\nonumber
\delta{\sigma}^{\mu\nu}_B&=&\frac{\tau_R}{12\pi^2}
\Bigg[\frac{10}{3}\mu\theta P^{\mu\nu}
-\Big(3\mu+2I_2T^3\tilde{U}_B\Big)\pi^{\mu\nu}\Bigg],
\end{eqnarray}
\begin{eqnarray}
\nonumber
\delta{\sigma}^{\mu\nu}_{\omega}&=&-\frac{\tau_R}{12\pi^2}
\Bigg[\frac{P^{\mu\nu}T^2\theta}{3}\Big(I_1+I_2T^2\tilde{U}_{\omega}\Big)
\\\nonumber
&&
+\pi^{\mu\nu}T^2\Big(\frac{20}{3}I_1-2I_2T^2\tilde{U}_{\omega}\Big)
\Bigg],
\end{eqnarray}
and
\begin{eqnarray}
\delta{\sigma}^{\mu\nu}_{\mathcal{A}}=-\frac{\tau_R\mu}{9\pi^2}\pi^{\mu\nu},
\end{eqnarray}
where $I_2=\bar{\mu}\big(\bar{\mu}^{2}+\pi^{2}\big)$. Note that there exist two terms in $\delta f^{I(Q)}_q$, $-\hbar\tau_R(2T)^{-1}(q\cdot \omega)(D T)(1-q_0\partial_{q_0})f^{(0)}_q$ and $-\hbar\tau_R(2q_0)^{-2}(q\cdot\partial)(q\cdot \omega)$, which also contribute to viscous corrections with vorticity when the hydrodynamic EOM are applied in computations.
In general, by redefining $u^{\mu}$, one can shift the above corrections on CME/CVE conductivities to the transport coefficients of energy density currents. 
It is more enlightening to simplify the expressions of $\delta\sigma^{\mu\nu}_B$ and $\delta\sigma^{\mu\nu}_{\omega}$ in distinct limits. In the high-temperature limit $(\bar{\mu}\ll 1)$, the coefficients in Eq.~\eqref{non_cond} reduce to
\begin{eqnarray}\nonumber
\delta\sigma^{\mu\nu}_B&=&\hbar\tau_R\mu\left(\frac{5P^{\mu\nu}}{18\pi^2}\theta-\frac{\pi^{\mu\nu}}{14\pi^2}\right),
\\\nonumber
\delta\sigma^{\mu\nu}_{\omega}&=&-\hbar\tau_R T^2\Bigg[\frac{P^{\mu\nu}}{108}\theta\left(1+\frac{111\bar{\mu}^2}{7\pi^2}\right)
\\
&&
-\frac{5\pi^{\mu\nu}}{27}\left(1-\frac{6\bar{\mu}^2}{7\pi^2}\right)\Bigg].
\end{eqnarray} 
On the contrary, in the low-temperature limit $(\bar{\mu}\gg 1)$, one obtains
\begin{eqnarray}\nonumber
\delta\sigma^{\mu\nu}_B&=&\hbar\tau_R\mu\left(\frac{5P^{\mu\nu}}{18\pi^2}\theta-\frac{\pi^{\mu\nu}}{6\bar{\mu}^2}\right),
\\
\delta\sigma^{\mu\nu}_{\omega}&=&-\hbar\tau_R\mu^2\left(\frac{P^{\mu\nu}}{12\pi^2}\theta
-\frac{2\pi^{\mu\nu}}{9\pi^2}\right).
\end{eqnarray} 

Despite the complexity of computations, it is worthwhile to note that such viscous corrections actually originate from Bianchi identities $\partial_{\nu}\tilde{F}^{\mu\nu}=\partial_{\nu}\tilde{\omega}^{\mu\nu}=0$, where $\tilde{\omega}^{\mu\nu}=\frac{1}{2}\epsilon^{\mu\nu\alpha\beta}\partial_{\alpha}u_{\beta}$, which relate the temporal derivatives of $B^{\mu}$ and $\omega^{\mu}$ (in the fluid co-moving frame) to their couplings with the gradients of $u^{\mu}$ as shown in Ref.~\cite{Hidaka:2017auj}. Such an origin is a reminiscence of the AC conductivity of CME driven by time-dependent $B^{\mu}$ and the viscous corrections upon CME/CVE are expected to be dissipative. 

So far, we have only considered the contributions for right-handed fermions. The quantum corrections for left-handed fermions will yield the same results but with the change of an overall sign for each term at $\mathcal{O}(\hbar)$.
Since $P^{\mu\nu}\theta$ and $\pi^{\mu\nu}$ are even under the parity ($\mathcal{P}$) transformation, it is anticipated that the transport coefficients of the non-equilibrium corrections on CME/CVE have the same parity as those in equilibrium, which can be more apparently observed from the simplified expressions shown in two limits above. Therefore, the bulk and shear strengths not only affect the vector currents ($J^{\mu}_V=J^{\mu}_R+J^{\mu}_L$) induced by CME/CVE with nonzero axial-charge chemical potentials ($\mu_A=\mu_R-\mu_L$) but also the axial currents ($J^{\mu}_A=J^{\mu}_R-J^{\mu}_L$) from the chiral separation effect(CSE) \footnote{Although CSE is known as a dual effect for CME in the vector/axial bases \cite{Fukushima:2008xe}, it is automatically included in the CME currents in the right/left-handed bases.} and CVE with nonzero vector-charge chemical potentials ($\mu_V=\mu_R+\mu_L$). It is worthwhile to note that such second-order quantum corrections on currents have different symmetry properties compared to the second-order classical effects also pertinent to magnetic fields such as the Hall-diffusion currents $J^{\mu}_{\perp}\sim \tau_R^2\epsilon^{\mu\nu\alpha\beta}u_{\nu}B_{\alpha}\partial_{\beta}\mu^2$ in Ref.~\cite{Gorbar:2016qfh}. Because there is no sign flipping for right/left-handed fermions in the classical case, the corresponding axial current can only be generated when $\mu_A\neq 0$(or $\partial_{\perp\beta}\mu_A\neq 0$), which is analogous to CESE \cite{Huang:2013iia,Pu:2014cwa}.

\section{Entropy production}
In contrast to the anomalous transport in equilibrium, the non-equilibrium quantum corrections of the charge current $\delta J_Q^{\mu}$ are dissipative. This is foreseen by the time-reversal symmetry ($\mathcal{T}$). Since the charge current $J_{\perp}^{\mu}$, $B^{\mu}$, and $\omega^{\mu}$ are $\mathcal{T}$-odd and $E^{\mu}$ is $\mathcal{T}$-even, from the classical ohmic current led by the RT approximation $J^{\mu}_{\perp}\propto \tau_R E^{\mu}$, one finds that $\tau_R$ is also $\mathcal{T}$-odd, whereas the CME/CVE conductivities in equilibrium are $\mathcal{T}$-even and thus non-dissipative. Nonetheless, because $\theta P^{\mu\nu}$ and $\pi^{\mu\nu}$ are both $\mathcal{T}$-odd, the corresponding transport coefficients of the viscous corrections upon chiral magnetic/vortical currents are proportional to $\tau_R$ multiplied by $\mathcal{T}$-even functions of $T$ and $\mu$, which accordingly yield dissipation. The same arguments can be applied to the non-equilibrium transport found in inviscid cases \cite{Hidaka:2017auj}. Notably, in light of symmetry, one finds $\mathcal{A}^{\mu}\sim B^{\mu}$ or $\mathcal{A}^{\mu}\sim \omega^{\mu}$ for viscous corrections. Consequently, the $\hbar$ corrections in practical collisions can only affect the prefactors of transport coefficients without altering their structures. Overall, the non-equilibrium second-order quantum transport is parity-odd and dissipative as opposed to the classical one (e.g., classical Hall effects), which is parity even and non-dissipative.

Albeit the non-equilibrium quantum transport results in dissipation, its entropy production is suppressed by the dissipation from classical effects as shown below. We may introduce the entropy density current in an usual form as for relativistic hydrodynamics, 
\begin{equation}
s^{\mu}=\frac{1}{T}(pu^{\mu}+T^{\mu\nu}u_{\nu}-\mu J^{\mu})
+\hbar (D_BB^{\mu}+ D_{\omega}\omega^{\mu}),
\end{equation}
where the non-dissipative corrections proportional to $B^{\mu}$ and $\omega^{\mu}$ originating from the non-dissipative charge and energy-density currents. The coefficients $D_{B}$ and $D_{\omega}$ should be determined by the transport coefficients of CME and CVE in equilibrium. See e.g., Ref.~\cite{Son:2009tf} for details. The explicit form of $D_{B(\omega)}$ is not important in our discussion. 

The constitutive relation now can be written as  $T^{\mu\nu}=u^{\mu}u^{\nu}\epsilon-pP^{\mu\nu}+\Pi^{\mu\nu}_{\text{dis}}+\Pi^{\mu\nu}_{\text{non}}$, where $\Pi^{\mu\nu}_{\text{dis}}$ denotes the dissipative corrections characterized by viscous effects and $\Pi^{\mu\nu}_{\text{non}}$ corresponds to the non-dissipative corrections led by anomalous effects in equilibrium. When $\mathcal{A}_{\mu}=0$, based on the matching conditions such that $u_{\mu}\delta J^{\mu}=u_{\mu}\delta T^{\mu\nu}=0$, we find
\begin{eqnarray}\label{DivS}
\partial_{\mu} s^{\mu}=\frac{1}{T}\Big[\Pi^{\mu\nu}_{\text{dis}}\partial_{\mu}u_{\nu}-(E_{\mu}+T\partial_{\mu}\bar{\mu})\delta J^{\mu}\Big].
\end{eqnarray}  
In Eq.~\eqref{DivS}, the classical contributions result in positive entropy production at $\mathcal{O}(\partial^2)$, while the corresponding quantum corrections are at $\mathcal{O}(\hbar\partial^3)$. Although the non-equilibrium quantum corrections here could be either positive or negative, they are always suppressed by the classical contributions and the second law of thermodynamics is satisfied.  

\section{Discussions and outlook}
Regarding the validity of our findings, due to the gradient expansion, the results should be legitimate for $\tau_R\partial\ll 1$, which imparts an upper bound for $\tau_R$. In addition, given that the CKT itself is subject to weakly coupled systems, $\tau_R$ cannot be too small. More generally, although the CKT is developed to cope with non-equilibrium conditions, it is applicable for small gradients and weak background fields based on the $\hbar$ expansion. To extend the validity beyond such limits, we have to solve for Wigner functions non-perturbatively from Kadanoff-Baym equations (derived from Schwinger-Dyson equations), which could be a formidable analytic problem. 

Nonetheless, we can still perturbatively solve for higher-order corrections in the $\hbar$ expansion (see Ref.~\cite{Gao:2018wmr} for a relevant study). Alternatively, one may also construct Wigner functions for non-equilibrium situations directly from Landau-level wave functions involving all-order $\hbar$ corrections, while it is subject to the case with just constant magnetic fields. Both approaches are systematic derivations of CKT from quantum field theory, which also complement each other. Up to $\mathcal{O}(\hbar^2)$, it is anticipated that novel non-dissipative quantum effects such as the charge modification led by $\omega\cdot B$ discovered in Ref.~\cite{Hattori:2016njk} will be found. Furthermore, for more general cases, one may include dynamical gauge fields, which could yield chiral-plasma instabilities \cite{Akamatsu:2013pjd} and more profound phenomena. 
%Such extensions and generalization should be pursued in the future. 
Such instabilities can be treated  in our hydrodynamics approach if the time and length scales of the instabilities are much larger than the  mean free time and length.
The typical  time and length scales of the instabilities are evaluated as  $\tau_{\text{inst}}\sim \tau_{R} m_{D}^{2}\ell_\text{inst}^{2}$ and $\ell_{\text{inst}}\sim 1/(\alpha\mu_A)$, where $\alpha=e^2/(4\pi)$ and $m_{D}$ is the Debye mass, while the mean free time and path are of order $\tau_{R}$~\cite{Akamatsu:2013pjd}.
Therefore, $\tau_\text{inst}\gg \tau_{R}$ and $\ell_\text{inst}\gg \tau_{R}$ lead to
conditions $m_{D}^{2}\gg (\alpha\mu_{A})^{2}$ and $\tau_{R}\ll 1/(\alpha\mu_{A})$. 
The former is satisfied at high temperature $T\gg \sqrt{\alpha}\mu_{A}$ or density $\mu_{V}\gg  \sqrt{\alpha}\mu_{A}$,
where $m_{D}^{2}\sim \alpha T^{2}$ or $\alpha\mu_{V}^{2}$. 
The latter depends on the strength of interaction. For example, if we evaluate $\tau_{R}$ in a quark gluon plasma, it behaves $\tau_{R}\sim \alpha_{s}T\ln\alpha_{s}^{-1}$, where $\alpha_{s}$ is the strong coupling constant; thus, the condition reduces to $T \gg \alpha\mu_{A}/(\alpha_{s}\ln\alpha_{s}^{-1})$.
If the opposite limit is realized, we have to employ the chiral kinetic equation to describe the dynamics of chiral kinetic instabilities.

%\red{In general, the approximated time scale for the  chiral-plasma instabilities obtained in Ref.~\cite{Akamatsu:2013pjd} is dictated by the axial-charge chemical potential, for example $\tau_{\text{inst}}\approx 1/(\alpha\mu_A)$ in electromagnetic plasmas, where $\alpha=e^2/(4\pi)$. The validity of hydrodynamics with an equilibrium $\mu_A$ requires $\tau_R\ll\tau_{\text{inst}}$. Although the magnitude of $\tau_R$ depends on the inverse strength of interactions between quasi-particles, one could generally approximate $\tau_R\sim 1/T$ by omitting the dependence in coupling based on hydrodynamics. It is thus found that our approach is legitimate when $\mu_5\ll T$ in most of cases.}

\acknowledgments
The authors thank I. Shovkovy for useful discussions.
Y. H. was partially supported by Japan Society of Promotion of Science (JSPS), Grants-in-Aid for Scientific Research
(KAKENHI) Grants No. 15H03652, 16K17716, and 17H06462.  Y. H. was also partially supported by RIKEN iTHES Project and iTHEMS Program.
D. Y. was supported by the RIKEN Foreign Postdoctoral Researcher program.

\newpage
\appendix

\section{Wigner functions with quantum corrections}\label{sec:WignerFunction}
Based on the Dirac equations under the Wigner transformation up to $\mathcal{O}(\hbar)$, we shall obtain the following Kaddanof-Baym-like equations for right-handed fermions,
\begin{eqnarray}
	&&\sigma^{\mu}\left(q_{\mu}+\frac{i\hbar}{2}\Delta_{\mu}\right)\grave{S}^{<}=\frac{i\hbar}{2}\left(\Sigma^<\grave{S}^{>}-\Sigma^>\grave{S}^{<}\right),
	\\
	&&\left(q_{\mu}-\frac{i\hbar}{2}\Delta_{\mu}\right)\grave{S}^{<}\sigma^{\mu}=-\frac{i\hbar}{2}\left(\grave{S}^{>}\Sigma^<-\grave{S}^{<}\Sigma^>\right),
\end{eqnarray}
By parameterizing $\grave{S}^<=\bar{\sigma}^{\mu}\grave{S}^<_{\mu}$, the above equation yield the difference equations,
\begin{eqnarray}\label{diff_Dirac}\nonumber
	&&
	\hbar\{\sigma^{\mu},\bar{\sigma}^{\nu}\}\mathcal{D}_{\mu}\grave{S}^<_{\nu}=2i[\sigma^{\mu},\bar{\sigma}^{\nu}]q_{\mu}\grave{S}^<_{\nu},
	\\
	&&
	\hbar[\sigma^{\mu},\bar{\sigma}^{\nu}]\mathcal{D}_{\mu}\grave{S}^<_{\nu}=2i\{\sigma^{\mu},\bar{\sigma}^{\nu}\}q_{\mu}\grave{S}^<_{\nu},
\end{eqnarray}
where $[A,B]=AB-BA$ and $\{A,B\}=AB+BA$ and
\begin{eqnarray}
	\mathcal{D}_{\mu}\grave{S}^{<}_{\nu}=\Delta_{\mu}\grave{S}^{<}_{\nu}-\Sigma^<_{\mu}\grave{S}^{>}_{\nu}+\Sigma^>_{\mu}\grave{S}^{<}_{\nu}
\end{eqnarray}
with $\Delta_{\mu}=\partial_{\mu}+F_{\nu\mu}\frac{\partial}{\partial q_{\nu}}$. We should now perturbatively solve for $\grave{S}^<_{\mu}$ from Eq.~\eqref{diff_Dirac} up to $\mathcal{O}(\hbar)$. We thus make the ansatz, $\grave{S}^<_{\mu}=\grave{S}^{<(0)}_{\mu}+\hbar \delta\grave{S}^<_{\mu}$. It is easy to find that
\begin{eqnarray}
	\grave{S}^{<(0)}_{\mu}=2\pi\delta(q^2)q_{\mu}f(q,X),
\end{eqnarray}  
which is independent of a choice for the basis of spins. We may now define $n\cdot\sigma=I$ by introducing a frame vector $n^{\mu}$ normalized as $n^2=1$ and its corresponding projection operator $P^{\mu\nu}=\eta^{\mu\nu}-n^{\mu}n^{\nu}$ giving $P^{\mu\nu}n_{\nu}=0$. The difference equations then become 
\begin{eqnarray}
	\mathcal{D}\cdot\grave{S}^<=0,\quad q\cdot\grave{S}^<=0,
\end{eqnarray}
and up to $\mathcal{O}(\hbar)$,
\begin{eqnarray}\label{traceless_part_cov}\nonumber
	&&2\pi\delta(q^2)\sigma^{\mu}_{\perp}\left(q\cdot n\mathcal{D}_{\mu}-q_{\mu}n\cdot\mathcal{D}\right)f=-2\sigma^{\mu}_{\perp}\epsilon_{\alpha\mu\nu\beta}n^{\alpha} q^{\nu}\delta\grave{S}^{<\beta},
	\\\nonumber
	&&
	2\pi\sigma^{\mu}_{\perp}\epsilon_{\alpha\mu\nu\beta}\delta(q^2)n^{\alpha}q^{\nu}\mathcal{D}^{\beta}f=2\sigma^{\mu}_{\perp}\left(q\cdot n\delta\grave{S}^<_{\mu}-q_{\mu}n\cdot\delta \grave{S}^<\right),
	\\
\end{eqnarray}
where $\sigma^{\mu}_{\perp}=P^{\mu\nu}\sigma_{\nu}$ and $\mathcal{D}_{\beta}f=\Delta_{\beta}f-\mathcal{C}_{\beta}$ and  $\mathcal{C}_{\beta}=\Sigma_{\beta}^<\bar{f}-\Sigma_{\beta}^>f$. When taking $n^{\mu}=(1,{\bf 0})$, Eq.~\eqref{traceless_part_cov} reduces to Eq.~(33) in Ref.~\cite{Hidaka:2016yjf}. Solving Eq.~\eqref{traceless_part_cov}, one derive
\begin{eqnarray}\nonumber
	\grave{S}^{<\mu}(q,X)
	&=&2\pi\Big(q^{\mu}\delta(q^2)
	+\hbar\delta(q^2)S_{(n)}^{\mu\nu}\mathcal{D}_{\nu}
	\\
	&&
	+\hbar\epsilon^{\mu\nu\alpha\beta}q_{\nu}F_{\alpha\beta}\frac{\partial\delta(q^2)}{2\partial q^2}
	\Big)f(q,X),
\end{eqnarray}
where 
\begin{eqnarray}
	S^{\mu\nu}_{(n)}=\frac{\epsilon^{\mu\nu\alpha\beta}}{2(q\cdot n)}q_{\alpha}n_{\beta} .
\end{eqnarray}
\\

\section{Frame transformation for the RTA}\label{sec:fRTA}
To preserve Lorentz invariance of the CKT in Eq.~\eqref{CKT} with the RTA, we have to introduce the frame transformation on $\mathcal{A}_{\mu}$. By implementing Eqs.~\eqref{CKT} and \eqref{RT_approx} can be rewritten as
	\begin{equation}\nonumber\label{CKT_RTA}
	\delta(q^2)\Big(\Box^{(n)}f^{(n)}_q+\frac{\hbar B^{(n)\mu}}{2q\cdot n}\partial_{q\mu}\big(q\cdot\Delta f^{(n)}_q\big)\Big)=I^{(n)}_c,
	\end{equation}
%	\begin{eqnarray}\nonumber\label{CKT_RTA}
%	&&\delta(q^2)\Bigg(\Box^{(n)}f^{(n)}_q+\frac{\hbar B^{(n)\mu}}{2q\cdot n}\partial_{q\mu}\Big(q\cdot\Delta f^{(n)}_q+\frac{q\cdot u}{\tau_R}\delta f^{(n)}_q\Big)\Bigg)
%	\\
%	&&=-\frac{\delta(q^2)}{\tau_R}\Big(q\cdot u+\frac{\hbar q\cdot\mathcal{A}^{(n)}}{(q\cdot u)^2}\Big)\delta f_q^{(n)},
%	\end{eqnarray}
with
	\begin{eqnarray}\notag
	I^{(n)}_c&=&-\frac{\delta(q^2)}{\tau_R}\Bigg(
	\Big(q\cdot u+\frac{\hbar q\cdot\mathcal{A}^{(n)}}{(q\cdot u)^2}\Big)\delta f^{(n)}_q 
	\\
	&&+\frac{\hbar B^{({n})\mu}}{2q\cdot n}\partial_{q\mu}\Big(q\cdot u\delta f^{(n)}_q\Big)
	\Bigg),
	\end{eqnarray}
	where we use the superindices $^{(n)}$ to track the frame-dependent terms explicitly. Also, we have applied the relations,
	\begin{eqnarray}
	\delta\Bigl(q^{2}-\hbar \frac{B^{(n)}\cdot q}{q\cdot n}\Bigr)=\delta(q^2)-\frac{\hbar B^{(n)\mu}}{2q\cdot n}\partial_{q\mu}\delta(q^2),
	\end{eqnarray} 
	and
	\begin{eqnarray}\nonumber
	(\partial_{q\mu}\delta(q^2))q\cdot{ \mathcal{D}}f^{(n)}_q=-\delta(q^2)\partial_{q\mu}(q\cdot\mathcal{D}f^{(n)}_q)+\mathcal{O}(\hbar).
	\\
	\end{eqnarray}
	Recall that the full CKT in Eq.~\eqref{CKT_RTA} is frame independent (Lorentz invariant) when considering also the nontrivial frame transformation upon the distribution functions.  
	According to Ref.~\cite{Hidaka:2016yjf}, the frame transformation for the distribution function reads
	\begin{eqnarray}\label{frame_trans_f}
	f^{(n')}_q=f^{(n)}_q+\hbar N^{\nu}_{nn'}\mathcal{D}_{\nu}f^{(n)}_q,
	\end{eqnarray}
	where 
	\begin{eqnarray}
	N^{\nu}_{nn'}=\frac{\epsilon^{\mu\nu\alpha\beta}q_{\alpha}n_{\beta}n'_{\mu}}{2(q\cdot n)(q\cdot n')}.
	\end{eqnarray}
	When taking $n^{\mu}=u^{\mu}$ and using the RTA, one finds that Eq.~\eqref{frame_trans_f} reduces to
	\begin{eqnarray}
	f^{(n')}_q=f^{(u)}_q+\hbar N^{\nu}_{un'}\Delta_{\nu}f^{(u)}_q,
	\end{eqnarray}
	which is independent of collisions. Note that the equation above also works for $\delta f^{(n')}_q$.
	Consequently, the collisional part alone in Eq.~\eqref{CKT_RTA} should be frame independent. 
%	We may extract such a part as
%	\begin{eqnarray}\nonumber
%	I^{(n')}_c&=&-\frac{\delta(q^2)}{\tau_R}\Bigg(
%	\Big(q\cdot u+\frac{\hbar q\cdot\mathcal{A}^{(n')}}{(q\cdot u)^2}\Big)\delta f^{(n')}_q
%	\\
%	&&+\frac{\hbar B^{(\blue{n'})\mu}}{2q\cdot n'}\partial_{q\mu}\Big(q\cdot u\delta f^{(n')}_q\Big)
%	\Bigg).
%	\end{eqnarray}
	By utilizing $I^{(n')}_c-I^{(u)}_c=0$ and $\delta f^{(n')}_q=\delta f^{(u)}_q+\hbar N^{\nu}_{un'}\Delta_{\nu}\delta f^{(u)}_q$, we obtain
	\begin{widetext}
		\begin{eqnarray}
				\mathcal{A}^{(n')}_{\nu}=\mathcal{A}^{(u)}_{\nu}-u_{\nu}(q\cdot u)^2\Bigg(N^{\alpha}_{un'}\Delta_{\alpha}+\frac{1}{2(q\cdot u)}\Big(\frac{B^{(n')\alpha}}{q\cdot n'}-\frac{B^{(u)\alpha}}{q\cdot u}\Big)\big(u_{\alpha}+(q\cdot u)\partial_{q\alpha}\big)\Bigg),
%		\mathcal{A}^{(n')}_{\nu}\delta f^{(n')}_q=\mathcal{A}^{(u)}_{\nu}\delta f^{(u)}_q-u_{\nu}(q\cdot u)^2\Bigg(N^{\alpha}_{un'}\Delta_{\alpha}+\frac{1}{2(q\cdot u)}\Big(\frac{B^{(n')\alpha}}{q\cdot n'}-\frac{B^{(u)\alpha}}{q\cdot u}\Big)\big(u_{\alpha}+(q\cdot u)\partial_{q\alpha}\big)\Bigg)\delta f^{(u)}_q,
		\end{eqnarray}
	\end{widetext}
	which gives the frame transformation upon the collisional terms in a RTA.
\\
	
\section{Divergence of the energy-momentum tensor}\label{sec:EM}
Here we present some critical steps for the derivation of Eq.~\eqref{div_Tmunu}. Following the trick in Ref.~\cite{Gorbar:2017toh}, we find
\begin{eqnarray}\nonumber
	\partial_{\mu}T^{\mu\nu}&=&\int\frac{d^4q}{(2\pi)^4}\Big(2q^{\nu}\partial\cdot S^<+q\cdot\partial S^{<\nu}-q^{\nu}\partial\cdot S^<\Big)
	\\\nonumber
	&=&\int\frac{d^4q}{(2\pi)^4}\Big(2q^{\nu}\partial\cdot S^<-\frac{1}{2}\epsilon^{\nu\kappa\sigma\rho}\epsilon_{\mu\lambda\sigma\rho}q^{\lambda}\partial_{\kappa}S^{<\mu}\Big).
	\\
\end{eqnarray}
Performing straightforward computations, one should obtain
%\begin{widetext}
	\begin{eqnarray}\nonumber
		\epsilon_{\mu\lambda\sigma\rho}q^{\lambda}\partial_{\kappa}S^{<\mu}
		&=&-\hbar\pi \bar{\epsilon}(q\cdot n)
		\Big[2\delta(q^2)q_{\sigma}\partial_{\kappa}\left(F_{\rho\lambda}\partial^{\lambda}_{q}f_q+\mathcal{C}_{\rho}\right)
		\\\nonumber
		&&
		+F_{\sigma\rho}q^{\lambda}(\partial_{q\lambda}\delta(q^2))(\partial_{\kappa}f_q)
		\\
		&&
		+2q_{\sigma}(\partial^{\lambda}_{q}\delta(q^2))\partial_{\kappa}\left(F_{\rho\lambda}f_q\right)
		\Big],
	\end{eqnarray}
%\end{widetext}
which results in
\begin{eqnarray}\nonumber\label{der_1}
	&&\int\frac{d^4q}{(2\pi)^4}(q\cdot\partial S^{<\nu}-q^{\nu}\partial\cdot S^<)\\
	&&=\hbar\int_q\delta(q^2)\frac{\epsilon^{\nu\kappa\sigma\rho}}{2}
	\Big[F_{\rho\sigma}\left(1+\frac{q^{\gamma}\partial_{q\gamma}}{2}\right)\partial_{\kappa}f_q
	+q_{\sigma}\partial_{\kappa}\mathcal{C}_{\rho}
	\Big]\notag
	\\
	&&=\hbar\int_q\delta(q^2)\frac{\epsilon^{\nu\kappa\sigma\rho}}{2}
	q_{\sigma}\partial_{\kappa}\mathcal{C}_{\rho},
\end{eqnarray}
where the first term on the right-hand side of the first equality in fact vanishes.
On the other hand, we find
\begin{eqnarray}\nonumber
	\int\frac{d^4q}{(2\pi)^4}2q^{\nu}\partial\cdot S^<
	&=&-2\int\frac{d^4q}{(2\pi)^4}q^{\nu}\Big(F_{\rho\mu}\partial_q^{\rho}S^{<\mu}
	\\
	&&-\Sigma^<\cdot S^>+\Sigma^>\cdot S^<
	\Big).
\end{eqnarray}
By performing the integration by part and dropping the divergent and vanishing surface terms, we obtain
\begin{widetext}
	\begin{eqnarray}\label{der_2}
		\int\frac{d^4q}{(2\pi)^4}2q^{\nu}\partial\cdot S^<
		&=&2\int\frac{d^4q}{(2\pi)^4}\Big[F^{\nu\mu}S^<_{\mu}+2\pi\bar{\epsilon}(q\cdot n)\delta(q^2)\Big(q^{\nu}q^{\mu}
		-\frac{\hbar}{4}q^{\nu}\epsilon^{\mu\sigma\alpha\beta}F_{\alpha\beta}\partial_{q\sigma}-\frac{\hbar}{4}\epsilon^{\mu\nu\alpha\beta}F_{\alpha\beta}\Big)\Big]\tilde{\mathcal{C}}_{\mu},
	\end{eqnarray}
\end{widetext}
up to $\mathcal{O}(\hbar)$.
Combining Eqs.~\eqref{der_1} and \eqref{der_2}, we acquire Eq.~\eqref{div_Tmunu}.

%\bibliography{CKT_ref}
\bibliography{viscous_CME_manuscript.bbl}
\end{document}